# TIME VARIATIONS OF THE SOLAR NEUTRINO FLUX DATA FROM SUDBURY NEUTRINO OBSERVATORY

Koushik Ghosh[1] and Probhas Raychaudhuri[2]

[1]**Department of Mathematics**
**Dr. B.C. Roy Engineering College**
**Durgapur-713 206**
**INDIA**
Email: koushikg123@yahoo.co.uk

[2]**Department of Applied Mathematics**
**University of Calcutta**
**92, A.P.C. Road, Calcutta-700 009**
**INDIA**
Email: probhasprc@rediffmail.com

## ABSTRACT

We have used the Ferraz-Mello Method of Date Compensated Discrete Fourier Transform on the $^8$B solar neutrino flux data from the Sudbury Neutrino Observatory (SNO) in order to search for its time variation. We present results from Ferraz-Mello Method for both pure $D_2O$ and salt phase data sets from SNO. For the $D_2O$ data the obtained periods are 5.36, 14.99, 23.06, 40.31, 65.63, 115.03, 131.27, 152.67, 188.74, 294.49 and 412.79 days and for the salt phase data the periods are obtained at around 37.84, 43.89, 59.17, 67.48, 72.47, 78.52, 86.54, 95.91, 109.83 and 317.01 days. For all the obtained periods the levels of confidence are sufficiently higher than 95%.

## I. INTRODUCTION

Solar neutrino flux detection is very important not only to understand the stellar evolution but also to understand the origin of the solar activity cycle. Recent solar neutrino flux observed by Superkamiokande [1] and SNO [2] suggest that solar neutrino flux from $^8$B neutrino and $^3$He+p neutrino from Standard Solar Model (S.S.M.) [3] is at best compatible with S.S.M. calculation if we consider the neutrino oscillation of M.S.W. [4] or if the neutrino flux from the sun is a mixture of two kinds of neutrino i.e. $\upsilon_e$ and $\upsilon_\mu$ [5]. S.S.M. is known to yield the stellar structure to a very good degree of precision but the S.S.M. cannot explain the solar activity cycle, the reason being that this S.S.M. does not include temperature and magnetic variability of the solar core [6, 7]. The temperature variability implied a variation of the energy source and from that source of energy magnetic field can be generated which also imply a magnetic variability [7]. The temperature variation is important for the time variation of the solar neutrino flux. So we need a perturbed solar model and it is outlined by Raychaudhuri since 1971 [6, 7] which may satisfy all the requirements of solar activity cycle with S.S.M. For the support of perturbed solar model we have demonstrated that solar neutrino flux data are fractal in nature [8, 9]. The excess nuclear energy from the perturbed nature of the solar model is transformed into magnetic energy, gravitational energy and thermal energy etc. below the tachocline. The variable nature of magnetic energy induces dynamic action for the generation of solar magnetic field.

Recently Yoo et al [10] searched the periodic modulations of the solar neutrino flux data of Superkamiokande-I (S.K.-I) detector from 31 May 1996 to 15 July 2001, almost half of the solar activity cycle; yielding a total detector life time of 1496 days. The solar neutrino flux data from S.K.-I, acquired for 1871 elapsed days from the beginning of data are divided into roughly 10-day-long samples as listed in Table I of the communication of Yoo et al [10]. It is observed that not all the data are perfectly of 10 days. They used Lomb Periodogram method for unevenly arranged sample data to search for possible periodicities in the S.K.-I solar neutrino flux data. They have found no statistical significance of the periodicities in the S.K.-I solar neutrino flux data. However, Caldwell and Sturrock [11] used almost the same method i.e. Lomb-Scargle method of analysis and they have found a very interesting period of 13.75 days in the solar neutrino flux data of S.K.-I apart from the other periods. In addition to this, there have been other reports of periodic variations in the measured solar neutrino fluxes from S.K.-I [12, 13]. The reported periods have been claimed to be related to the solar rotational period. Particularly there is claim [12, 13] of 7% amplitude modulation in S.K.-I neutrino flux data at a frequency of 9.43 year$^{-1}$. Thus there arises a controversy regarding the periodicities of the solar neutrino flux data. We have analyzed the time variations of the solar neutrino flux data from S.K.-I by Ferraz-Mello Method, Periodogram Method and Rayleigh Power Spectrum Analysis [14, 15] and found distinct periods for the data. Apart from the S.K.-I data we have also analyzed the time variations of SAGE and GALLEX-GNO data by Ferraz-Mello Method, Periodogram Method and Rayleigh Power Spectrum Analysis [16, 17] and found conspicuous periods for them also.

Aharmim et al [18] presented a search for periodicities in the SNO's pure $D_2O$ and salt phase data sets by an unbinned Maximum Likelihood Analysis and Lomb-Scargle Periodogram Analysis. The $D_2O$ data set consists of 559 runs starting on November 2, 1999 and spans a calendar period of 572.2 days during which the total neutrino life-time was 312.9 days. The salt phase of SNO started on July 26, 2001, 59.7 calendar days after the end of the pure $D_2O$ phase of the experiment. The salt data set contains 1212 runs and spans a calendar period of 762.7 days during which the total neutrino life-time was 398.6 days. Aharmim et al [18] found no significant sinusoidal periodicities with periods between 1 day and 10 years with either an unbinned Maximum Likelihood Analysis or by Lomb-Scargle Periodogram analysis. Ranucci and Rovere also in their very recent communication [19] did not get any hint of periodicities for SNO data within the sensitivity limits by standard Lomb-Scargle Method and by a Likelihood Extension of Lomb-Scargle Method.

In the present paper in order to analyze the SNO's pure $D_2O$ and salt phase data sets we first make 3-point moving average of these data to reduce the noise present in the data and then apply Ferraz-Mello Method of Date Compensated Discrete Fourier Transform in the processed data in order to search for their time variation. The observation of a variable nature of solar neutrino would provide significance to our understanding of solar internal dynamics and probably to the requirement of the modification of the Standard Solar Model i.e. a perturbed solar model.

## II. CALCULATION OF PERIODICITIES BY FERRAZ-MELLO METHOD OF DATE-COMPENSATED DISCRETE FOURIER TRANSFORM

The technique called Date-Compensated Discrete Fourier Transform (DCDFT) corresponds to a curve-fitting approach using a sinusoid-plus-constant model and summarized below. For each trial frequency $\omega$, one coefficient of spectral correlation S is obtained by the following formulae [20]:

$$a_0^{-2} = N \tag{1}$$

$$a_1^{-2} = \Sigma \cos^2 x_i - a_0^2 (\Sigma \cos x_i)^2 \tag{2}$$

$$a_2^{-2} = \Sigma \sin^2 x_i - a_0^2 (\Sigma \sin x_i)^2 - a_1^2 M^2 \tag{3}$$

where

$$M = \Sigma \cos x_i \sin x_i - a_0^2 (\Sigma \sin x_i)(\Sigma \cos x_i) \tag{4}$$

and

$$c_1 = a_1 \Sigma f_i \cos x_i \tag{5}$$

$$c_2 = a_2 \Sigma f_i \sin x_i - a_1 a_2 c_1 M \tag{6}$$

$$S = (c_1^2 + c_2^2) / \Sigma f_i^2 \tag{7}$$

N is the number of observations in the series, $t_i$ are the observation dates, $x_i = 2\pi\omega t_i$ and $f_i$ are the measures of data referred to the mean i.e. $f_i = y_i - \overline{y}$ so that $\Sigma f_i = 0$ where $y_i$'s are the observed data. All other symbols are intrinsic quantities. The summations are made for i=1 to i=N. Usually the range of frequencies is considered from the observed frequency $\omega_{obs} = 1/N$ to the Nyquist frequency $\omega_N$.

To decide whether the peaks in the graph are significant or not we use the test derived by G.R. Quast [21] which is given by the following expressions:

$$G = -(N-3) \ln(1-S)/2 \tag{8}$$

$$H = (N-4)(G + e^{-G} - 1)/(N-3) \tag{9}$$

$$\alpha = 2(N-3) \Delta t \, \Delta\omega / 3(N-4) \tag{10}$$

$$C = (1 - e^{-H})^\alpha \tag{11}$$

where $\Delta t$ is the time interval covered by the observations and $\Delta\omega$ is the range of frequencies sampled. C is the confidence of the result; (1-C) may be interpreted as the probability of having the height of the highest peak by chance only. We consider a suitable number of values of $\omega$ from $\omega_{obs} = 1/N$ to $\omega_N$ and we arrange the corresponding days from $1/\omega_N$ to $1/\omega_{obs}$ at equal intervals and we obtain corresponding magnitudes of H and C at those days. Next we plot the H-$\omega$ graph for the considered data.

## III. RESULTS

Before to start processing of data by Ferraz-Mello Method we make the 3-point moving average of both pure $D_2O$ and salt phase data sets to reduce the noise present in the data. The processed data sets are used for further analysis. In the present analysis we consider 5000 values of ω from $\omega_{obs}$ to $\omega_N$ to make the steps sufficiently small to avoid missing of an important peak for both pure $D_2O$ and salt phase data sets. For both the data our choice of Nyquist period is 2 days i.e. the Nyquist frequency is 0.5 day$^{-1}$. The results are listed below:

| DATA | PERIODS (IN DAYS) OBTAINED BY FERRAZ-MELLO METHOD OF DATE COMPENSATED DISCRETE FOURIER TRANSFORM [THE CORRESPONDING LEVELS OF CONFIDENCE ARE GIVEN WITHIN BRACKETS] |
|---|---|
| SNO's $D_2O$ data | 5.36 (95.94%), 14.99 (96.52%), 23.06 (97.90%), 40.31 (98.89%), 65.63 (97.81%), 115.03 (99.44%), 131.27 (98.73%), 152.67 (97.67%), 188.74 (99.85%), 294.49 (99.77%), 412.79 (99.99%). |
| SNO's salt phase data | 37.84 (99.75%), 43.89 (96.28%), 59.17 (99.08%), 67.48 (99.26%), 72.47 (99.57%), 78.52 (99.69%), 86.54 (99.79%), 95.91 (99.99%), 109.83 (99.97%), 317.01 (99.99%). |

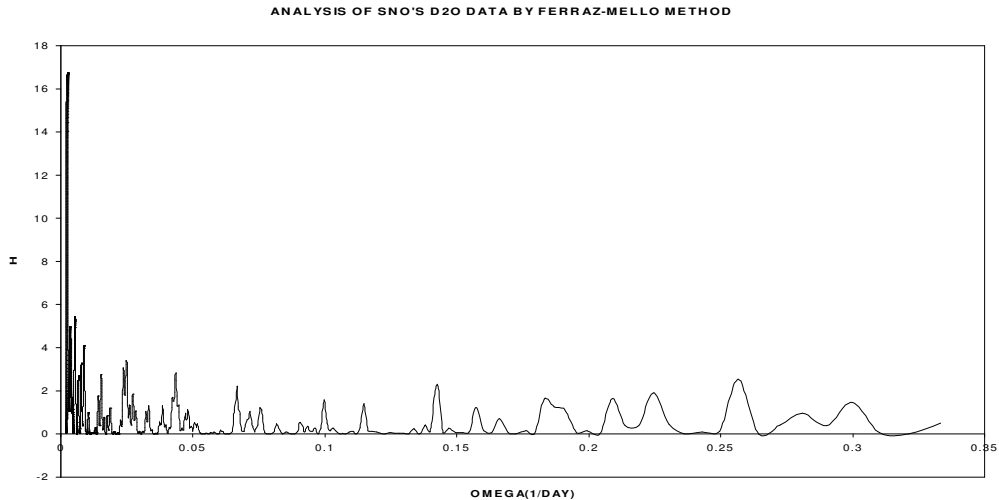

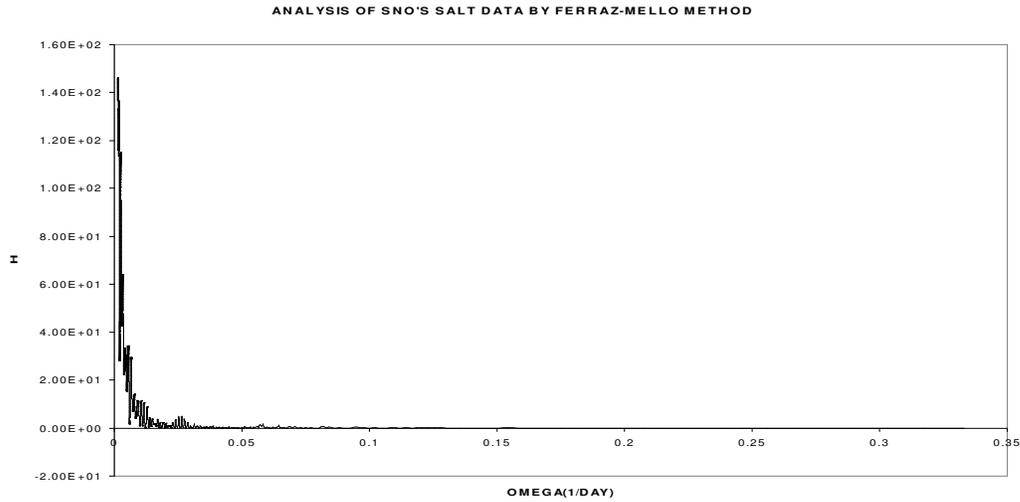

## IV. DISCUSSION

J. Yoo et al. [10] in their analysis of S.K.-I data have taken the raw data for consideration although noise is very much present in S.K.-I data. B. Aharmim et al. [18], Ranucci and Rovere [19] also considered the raw data from SNO for their discussions without removing the noise from the data. This kind of treatment with the presence of noise in the data can often fail to give a proper analysis of the data. We can certainly expect a better result from our method since we have gone through the initial process to smooth the data by applying 3-point moving average on it. The observed period of 14.99 days for the $D_2O$ data is not appreciably different from the period of 13.75 days for the S.K.-I data obtained by Caldwell and Sturrock [11]. The period of 40.31 days for $D_2O$ data and the periods of 37.84 and 43.89 days for the salt phase data are appreciably similar to the period corresponding to the frequency of 9.43 year$^{-1}$ for the S.K.-I data obtained by Sturrock [12, 13]. We here get a period of 5.36 days of $D_2O$ data which is similar to the period of 0.25 months obtained for 5-days-long S.K.-I data by Rayleigh Power Spectrum Analysis [15]. The observed period of 14.99 days for the $D_2O$ data is very much similar to the period of 0.5 months for 10-days-long S.K.-I data by Rayleigh Power Spectrum Analysis [15]. The obtained period of 40.31 days for $D_2O$ data and the periods of 37.84 and 43.89 days for the salt phase data are appreciably similar to the period of 1.31 months obtained for 10-days-long S.K.-I data by Ferraz-Mello Method [14]. Again the detected peak at 412.79 days is almost similar to the obtained peak at 14.01 months for 45-days-long S.K.-I data by Ferraz-Mello Method [14].